# Simplified Klinokinesis using Spiking Neural Networks for Resource-Constrained Navigation on the Neuromorphic Processor Loihi

Apoorv Kishore
*Dept. of Electrical Engineering,*
IIT Bombay
Mumbai, India

Vivek Saraswat
*Dept. of Electrical Engineering*
IIT Bombay
Mumbai, India

Udayan Ganguly
*Dept. of Electrical Engineering*
IIT Bombay
Mumbai, India

*Abstract*— C. elegans shows chemotaxis using klinokinesis where the worm senses the concentration based on a single concentration sensor to compute the concentration gradient to perform foraging through gradient ascent/descent towards the target concentration followed by contour tracking. The biomimetic implementation requires complex neurons with multiple ion channel dynamics as well as interneurons for control. While this is a key capability of autonomous robots, its implementation on energy-efficient neuromorphic hardware like Intel's Loihi requires adaptation of the network to hardware-specific constraints, which has not been achieved. In this paper, we demonstrate the adaptation of chemotaxis based on klinokinesis to Loihi by implementing necessary neuronal dynamics with only LIF neurons as well as a complete spike-based implementation of all functions e.g. Heaviside function and subtractions. Our results show that Loihi implementation is equivalent to the software counterpart on Python in terms of performance - both during foraging and contour tracking. The Loihi results are also resilient in noisy environments. Thus, we demonstrate a successful adaptation of chemotaxis on Loihi - which can now be combined with the rich array of SNN blocks for SNN based complex robotic control.

## I. INTRODUCTION

Caenorhabditis elegans is a transparent multicellular nematode that lives in a soil environment and has been extensively used for neural research over the years by virtue of its transparency. Studies involving C.elegans constitute phenomena like thermotaxis, chemotaxis, learning, and memory [1]. The phenomenon of interest here is chemotaxis: the ability of the organism to sense and remember salt concentrations that it has experienced under starvation conditions and to move towards a more desirable salt concentration. This ability to forage for food and track the desired concentration is known as contour tracking and it has been experimentally observed in the worm [1]. Contour tracking is performed by the worm employing various mechanisms like Klinokinesis, Klinotaxis, and Orthokinesis, however, the focus of this paper will be restricted only to Klinokinesis in which the worm makes abrupt turns away from its current direction to navigate. In

| | | Santurkar and Rajendran 2015 | This paper 2021 |
|---|---|---|---|
| Properties | Spiking Domain | ✓ | ✓ |
| | Autonomous | ✓ | ✓ |
| Network Architecture | Neuron Models | ASE Neuron $\tau_m \dot{V} = (V_0 - V) + k^d(V_d - V) + k^h(V_h - V)$ $\alpha_L^d = \alpha_{L0}^d(C - C_L) \times H(C - C_L)$ $\alpha_R^d = \alpha_{R0}^d(C - C_R) \times H(C_R - C)$ $\dot{C}_L = (C \times H(C - C_L) - C_L)/\tau_L$ $\dot{C}_R = (C \times H(C_R - C) - sgn(C_R - C)C_R)/\tau_R$ LIF Neuron $\dot{I} = \frac{-I}{\tau_1} + \sum_j w_{ij} \cdot s_j(t)$  $\dot{V} = \frac{-V}{\tau_2} + I + I_{bias}$ | LIF Neuron $\dot{I} = \frac{-I}{\tau_1} + \sum_j w_{ij} \cdot s_j(t)$ $\dot{V} = \frac{-V}{\tau_2} + I + I_{bias}$ |
| | Absence of external calculations | ✗ | ✓ |
| Neuromorphic Hardware Implementable | Loihi [Neuron modelling: LIF] | ✗ | ✓ |

*Table.{1}. Comparison of the paper with the past literature*

Klinokinesis, the worm senses the present concentration and compares it with the immediate past concentration to move. It changes the course of its motion when it senses that it is moving in an undesirable direction. It deviates from the current path if it senses that it is above(below) the setpoint concentration and that the present gradient is positive(negative) and thus corrects itself constantly to ensure its movement is aligned with the target. A circuit that replicates the function of left and right amphid sensory neurons of C.elegans was introduced in [2]. Santurkar and Rajendran further proposed an algorithm designed for a Spiking Neural Network (SNN) which employs motor neurons and amphid sensory neurons to perform klinokinesis [3]. They also demonstrate the energy efficiency of the SNN based chemotaxis and contour tracking [3].

Further studies examining chemotaxis illustrate the development of a complex XOR function using SNNs to create neural circuits for escaping local minima [4] as well as a more efficient combination of Klinokinesis, Klinotaxis, and Orthokinesis for contour tracking [5]. Thus, comprehensive navigational functionalities have been demonstrated in software.

Dedicated SNN hardware further enables the energy-efficient operation to capacitate autonomous navigation robots that function under grueling resource constraints (neuronal complexity and network size) and consequently energy constraints. Dedicated ASIC with nanoscale neurons [4] can lead to custom design [6]. However, standard SNN like Intel's neuromorphic processor codenamed Loihi provides a common choice of neurons and synapse behaviors. Hence, hardware implementation requires algorithmic adaptation to utilize the limited hardware capabilities as opposed to general-purpose CPU functions. Once implemented, the navigational SNNs can be run efficiently. Additionally, these networks may be coupled with other navigational elements like Simultaneous Localization And Mapping (SLAM) implemented on Loihi [7]. Loihi is the first fully integrated SNN chip that supports on-chip learning, multi-core connectivity, multilayer dendritic trees, and reward-based learning which provides a standard neural processor interface for highly efficient computing [8].

The primary aim of this paper is to implement chemotaxis-based contour tracking using klinokinesis on Loihi. Santurkar & Rajendran employ a combination of Leaky-Integrate and Fire (LIF) neurons as well as biologically inspired Amphid sensory (ASE) neurons that are not hardware friendly due to their underlying complex update equations and parameter requirements. The ASE neuron model has eight state variables including transition rates of ion channels, membrane potential, and internal threshold concentrations which are used to sense the present concentration gradient. In contrast, Loihi supports the Leaky-Integrate and Fire neuron model which has only two internal state variables, namely membrane current and potential [8]. Hence, the challenge is to adapt the original SNN in [3] to simple LIF neurons to demonstrate hardware implementation. In this paper, we relieve the requirement of ASE neurons by performing gradient detection using simple LIF neurons with two state variables, hence making the algorithm hardware friendly. Further, all non-spiking signal processing operations used in [3] like the Heaviside function H(.), concentration subtraction, etc. are implemented on Loihi using spikes. Thus, a completely spike-based klinokinesis-based chemotaxis is designed and implemented on Loihi.

In the subsequent sections of this paper, we discuss our proposed network architecture along with the functionalities of individual blocks in the network. We further discuss hardware simulations of the algorithm in the presence/absence of noise and the striking concordance in the hardware & software simulation results of the algorithm which speaks to the fidelity of implementation on Loihi.

## II. NETWORK ARCHITECTURE

### A. Klinokinesis with Single Concentration Sensor

C. Elegans employs the information of its immediate previous position from memory to steer itself in the desired direction. The end goal could be to reach a preferable concentration of NaCl in the environment which is correlated with its starvation. This desired end concentration is called the setpoint. The navigation process can be modeled through mechanisms like Klinokinesis, Klinotaxis, and Orthokinesis.

Klinokinesis uses the single concentration sensor in the worm to regularly sense the current concentration (C) and utilize its memory of the past concentration to estimate the concentration gradient in its path and navigate towards the desired setpoint concentration ($C_T$). The essential aspects of the algorithm are as follows: If the worm is far away from $C_T$ and cannot detect any surrounding gradient, it traverses its neighborhood randomly in search of food until it encounters a non-zero gradient. If the worm senses the gradient to be positive(negative), and the present C < $C_T$(C > $C_T$), then it assumes the present direction of motion as favorable otherwise, it changes the course of its motion to subsequently align itself with the correct gradient. It is important to note that the klinokinesis uses only the sign of the gradient. In the following subsections, we discuss the network used to implement this algorithm in Loihi.

### B. Neuron Model

The algorithm is implemented using the Leaky-Integrate and Fire (LIF) neuron model which is the neuron model supported by Loihi [8]. The membrane current and potential update equations of a typical LIF model used in Loihi are as follows:

$$u_i(t) = u_i(t-1) \cdot (2^{12} - \delta_i^{(u)}) \cdot 2^{-12} + 2^6 \sum_j w_{ij} \cdot s_j(t)$$
$$v_i(t) = v_i(t-1) \cdot (2^{12} - \delta_i^{(v)}) \cdot 2^{-12} + u_i(t) + u_{bias}(t)$$

where $u_i(t)$ and $v_i(t)$ denote the membrane current and potential of the $i^{th}$ neuron at time t respectively. $\delta_i^{(u)}$ and $\delta_i^{(v)}$ represent the current and voltage decay. $w_{ij}$ is the weight of the synapse originating from the $j^{th}$ neuron and connecting the $i^{th}$ neuron. $s_j(t)$ holds value 1 if the $j^{th}$ neuron has spiked at time t and holds zero otherwise. $u_{bias}(t)$ is the bias current or external input current to the neuron at time t. The $i^{th}$ neuron produces a spike when its potential $v_i(t)$ exceeds a certain potential threshold $v_{TH}$, post which, its potential is reset back to zero. The various parameter and their corresponding values used in our simulations are tabulated below:

| Parameter | Value (unit-less) |
|---|---|
| $\delta_i^{(u)}$ | $2^{12}$ |
| $\delta_i^{(v)}$ | 1 |
| $w_{excitatory}$ | 128 |
| $w_{inhibitory}$ | $-128$ |
| $u_{bias}$ | 0 |
| $v_{TH}$ | 6400 |

*Table.{2}. LIF neuron model parameters for simulations*

### C. Working Principle

A favorable direction of further motion for the worm is determined by finding out whether the worm is currently moving towards an increasing or decreasing concentration value.

Hence, determining whether concentration is currently increasing(gradient > 0) or decreasing(gradient<0) is of primary concern in klinokinesis. As mentioned in Section I, [3] achieves this functionality of estimating the sign of the gradient by employing ASE neurons. ASER neurons take as input $H(C-C_R)$ or $H(C_R-C)$, where $C_R$ ($C_L$ used for ASEL neurons) is an internal neuron parameter representing the previous concentration and $H(.)$ is the heaviside or unit step function. In this algorithm, however, we use the heaviside function implemented in spikes to determine the sign of the gradient. Consider $C(t)$ as the present concentration of the worm and $C(t-1)$ as the concentration of the immediate previous position. If

$$H(C(t) - C(t-1)) = 1, \ H(C(t-1) - C(t)) = 0$$

Then
$$C(t) - C(t-1) > 0$$

implying that the gradient > 0. If

$$H(C(t-1) - C(t)) = 1, \ H(C(t) - C(t-1)) = 0$$

Then
$$C(t-1) - C(t) > 0$$

which implies that the gradient < 0. However, If both the Heaviside functions are zero, then it simply means that $C(t)= C(t-1)$ and the gradient is zero as well.

Fig.{1} demonstrates how we implement the Heaviside function using spiking neurons. The concentration values are frequency encoded in spikes such that the higher the concentration, the higher is the spike rate. Due to the spike encoding of concentration values, temporal spike patterns cannot reveal instantaneous results but require a time frame to perform subtraction during which the concentrations remain fixed as can be observed from the figure. This pre-defined time frame is referred to in this paper as the time window demonstrated by the vertical black dashed lines in the figure and has a fixed value of 4000 time steps for all the simulations. If the resultant neuron spikes more than once in the corresponding time window, then the result of the Heaviside is said to be 1, else it is 0. In the figure, $C(t)$ is represented as $C_1$, and $C(t-1)$ is represented as $C_2$. It can be observed from the figure that in the first time window, $C_1 < C_2$, which means that $H(C_1 - C_2)$ [represented by $N_3$] should be 0 and $H(C_2 - C_1)$ [represented by $N_4$] should be 1 which agrees with the spiking approximation of result. It can also be observed from the response of $N_3$ that the subtraction operation is capable of showing the correct result only if operated over a long enough time window. Another important thing to note is that at the beginning of the third time window, $N_3$ spikes once due to residual potential but the Heaviside still shows the correct result which demonstrates the robustness of the algorithm.

### D. Concentration Sensor

The concentration sensor is implemented using the interactive spike sender-receiver functionality available in Loihi. The concentration sensing and encoding are executed as follows: The concentration space is generated as a two-dimensional matrix with each matrix element representing a point in 2D space and specifying a concentration value. Whenever the worm is at a position (x,y), the concentration value at that point is read and passed on to the interactive spike sender which in turn encodes that concentration value in spikes and sends it to the network. This operation is performed throughout every time window to sense the present concentration since the interactive spike sender sends spikes in real-time with the maximum possible frequency of 100Hz.

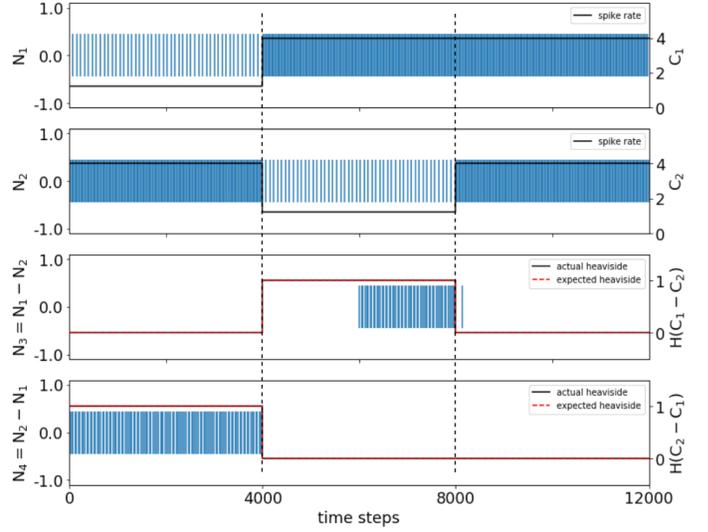

*Fig.{1}. A Spiking implementation of the Heaviside function. Neuron $N_1$ encodes concentration $C_1$ and N2 encodes concentration $C_2$ (spike rate is represented by the black solid line). $C_1$ and $C_2$ are subtracted by $N_3$ and $N_4$ to replicate the functionality of $H(C_1 - C_2)$ and $H(C_2 - C_1)$ respectively. The expected and actual results are represented by the red dashed and black solid lines respectively which are coincident in the final two plots. The vertical black dashed lines indicate the duration of a time window*

This frequency is thus the theoretical upper limit of our concentration encoding. The fact that subtraction in the spiking domain is only an approximation further limits the concentration values that we can encode because if the spike rates are too close for $C_1$ and $C_2$, then $C_1 - C_2$ shows little to no response in our allotted time window. Then, the window would have to be increased which cannot be afforded at the expense of run-time. Our concentration space hence has limited and quantized concentration values that are suitable for sensing and spike operations.

### E. Gradient Detection

The block diagram of the proposed network is shown in Fig.{2}. The present concentration sensed from the interactive spike sender acts as the response of the neuron $N_{present}$ which when delayed by a time window length becomes the spike train encoding of the previous concentration and hence the response of neuron $N_{prev}$.

Similarly, the setpoint concentration is generated from a basic spike sender for neuron $N_{set}$. Just as demonstrated in subsection B, the responses of the neurons $N_{gradpos}$ and $N_{gradneg}$ are the approximation to $H(C(t) - C(t-1))$ and $H(C(t-1) - C(t))$ which in turn determine the sign of the gradient in the present time window as positive and negative respectively.

## F. Navigation

From the block diagram shown in Fig.{2}, it can be observed that the neurons $N_{high}$ and $N_{low}$ respond when the present concentration $C(t)$ is higher and lower than the setpoint $C_T$ respectively. If $C(t) > C_T$ then it implies that the worm must align itself towards a negative gradient or If $C(t) < C_T$ then it must align itself towards a positive gradient to reach the setpoint. If both $C(t)$ and $C_T$ hold the same value, then it must mean that the worm has reached its destination successfully and hence must stay around it. In our algorithm, there is no concept of velocity since the concentration space is quantized and hence the worm changes its position by 1 element of the 2D space matrix after each time window. Since the concentration space is also quantized, there might occur instances where a plateau (zero gradient region) is encountered near the setpoint concentration. In such cases, the worm is tuned to perform a random walk after which it would reach the edge of that plateau and realign itself towards the correct direction.

The navigation control layer works as follows: If $C(t) < C_T$ ($N_{low}$ is spiking) and the worm is moving on a negative gradient ($N_{gradneg}$ is spiking), then this direction of motion is undesirable as the worm must rise in terms of surrounding concentration, in this case, the worm is tuned to turn anti-clockwise by 45°. Similarly, If $C(t) > C_T$ ($N_{high}$ is spiking) and the worm is moving on a positive gradient ($N_{gradpos}$ is spiking), then the direction is undesirable and it is tuned to turn clockwise by 45°. Finally, if neither $N_{gradpos}$ nor $N_{gradneg}$ are spiking then the worm is tuned to explore randomly by choosing an angle of turning with equal probability from {-45°, 0°, 45°}. This can be modeled by providing a bias current to the random walk neuron so that it also captures the case in which it encounters a plateau at the setpoint concentration, i.e. none of the 4 neurons from the second layer are spiking, and hence, in this case, as well, the worm will perform random exploration.

For the rest of the possible cases, in which the present direction is desirable, the direction of the motion of the worm is kept unaltered. The magnitude of the turning angle is not continuous and fixed to 45° because the concentration space is not continuous and hence rotation is possible only in multiples of 45°. For the subsequent simulations in this paper, the final layer neurons have not been explicitly implemented as neurons in Loihi, rather, only the functionality has been captured for the sake of time.

## G. Network Response

Fig.{3} demonstrates a sample temporal concentration profile of the worm and the corresponding responses of each neuron in the network. The first plot

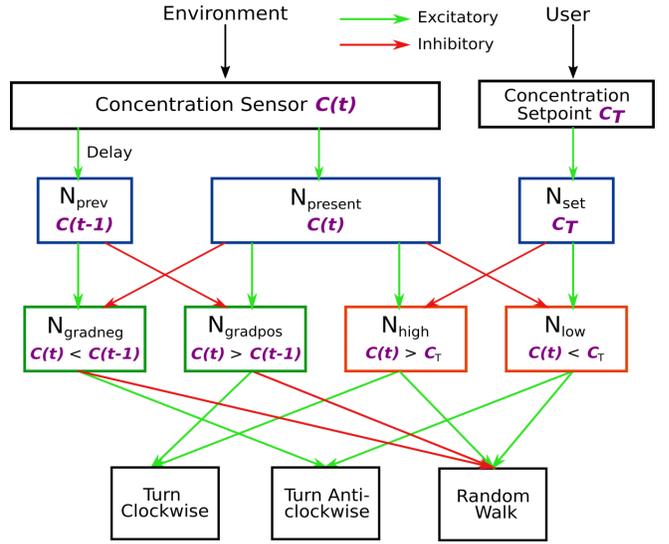

*Fig.{2}. The Block diagram for the hardware implementable SNN klinokinesis algorithm that takes the worm from an initial concentration of NaCl to the desired setpoint $C_T$. Neurons in the first layer ($N_{present}$, $N_{prev}$, and $N_{set}$) encode the corresponding concentrations in spikes (present, previous, and setpoint). Neurons $N_{gradneg}/N_{gradpos}$ determine the sign of the gradient. Neurons $N_{high}/N_{low}$ determine whether it has to ascend or descend the gradient. The neurons in the final layer determine the further course of motion*

shows the temporal concentration profile, the setpoint concentration, and the point of maxima. It also divides the concentration space into 4 overlapping regions where Gradient > 0 (region before the maxima), Gradient < 0 (region after the maxima), $C(t) < C_T$ (region before the point of intersection of C & $C_T$), and $C(t) > C_T$ (region after the point of intersection of C & $C_T$) are pointed out. These regions are color-coded the same as the neurons whose spiking behavior determines whether the condition of that region is true. The second plot shows the encoding of chosen concentration values by $N_{present}$ for each time window. The third and fourth plots show the delayed response of $N_{prev}$ and the constant setpoint response of $N_{set}$. The fifth and sixth plots validate the coinciding of the spiking regions of $N_{gradneg}$ and $N_{gradpos}$ with the Gradient < 0 and Gradient > 0 regions (green) marked in the first plot respectively. The final two plots also validate the coinciding of spiking regions of $N_{high}$ and $N_{low}$ with the $C(t) > C_T$ and $C(t) < C_T$ regions (red) marked in the first plot respectively. It can be noticed that there are time windows in which neither $N_{high}$ nor $N_{low}$ are spiking. In these time windows, the present concentration is equal to the setpoint. It should be noted that this is just a sample profile meant only for studying the neuron responses and hence navigational rules are not enforced here.

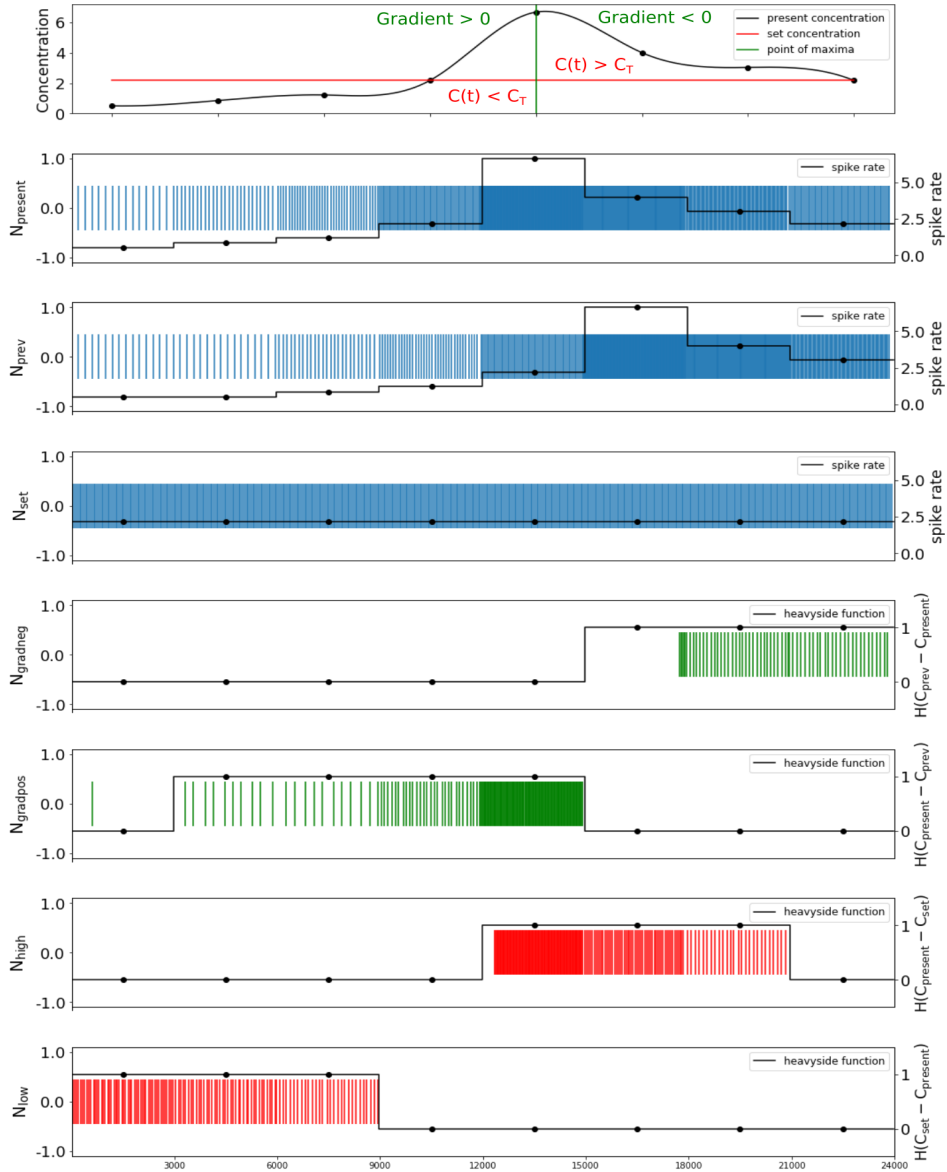

*Fig.{3}. A sample concentration profile and the corresponding neuron responses with their spike rates. The topmost plot displays the sample concentration, the setpoint, the point of maxima, and the regions marking positive/negative gradient and higher/lower concentration than the setpoint. The spike rate of $N_{present}$ is proportional to the concentration profile (second plot). $N_{prev}$ has the delayed response of $N_{present}$ signifying the previous position (third plot). $N_{set}$ represents the setpoint concentration which is temporally constant (fourth plot). $N_{gradneg}(N_{gradpos})$ displays the response in the region where the gradient is negative(positive) (fifth and sixth plot). $N_{high}(N_{low})$ displays the response in the region where the present concentration is higher(lower) than the setpoint (seventh and eighth plot)*

### III. SIMULATION AND RESULTS

The simulations are set in a concentration space of size 101 x 101 where each dimension spans a range [0,100]. The concentration space is quantized using 20 concentration values given by {0.25Hz, 0.37Hz, 0.5Hz, 0.67Hz, 0.85Hz, 1Hz, 1.1Hz, 1.22Hz, 1.39Hz, 1.61Hz, 1.89Hz, 2.17Hz, 2.56Hz, 3.03Hz, 3.45Hz, 4Hz, 4.76Hz, 5.56Hz, 6.67Hz, 8.33Hz} carefully chosen such that each concentration is easily distinguishable from another in spiking domain and the subtraction operation yields a result in an affordable time window of duration 4000 time steps. The setpoint concentration for all the simulations is fixed to $C_T = 4.0$Hz.

Fig.{4} is a typical simulation demonstrating the ability of the worm-bot to perform a random exploration

when it is unable to sense any gradient. The worm is launched at the point (10,50) [purple dot] with concentration value C = 0.25Hz. Towards the end of the simulation, it starts to experience a change in the gradient sign and the trajectory hence starts to become more aligned.

$\sigma = 0.05$ is plotted in Fig.{8}. The average absolute deviation during the tracking of the contour for a noisy trajectory was found to be ~7.08 units in the 101x101 space which underlines the robustness of the algorithm.

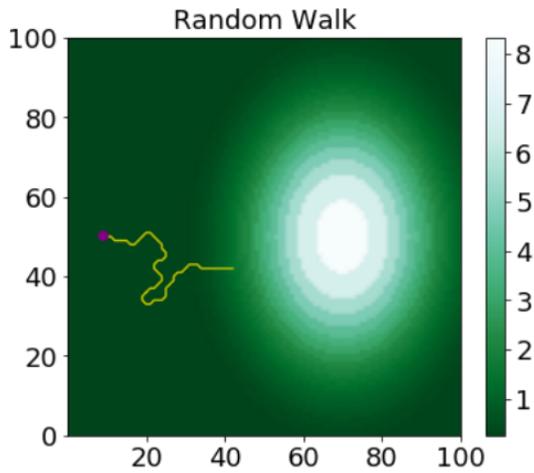

Fig.{4}. *The worm bot performs random walk initiating from the point (10,50) [purple dot] with C = 0.25Hz and the specified $C_T$ = 4.0Hz. On encountering a positive gradient, the trajectory starts to evolve in a straight line*

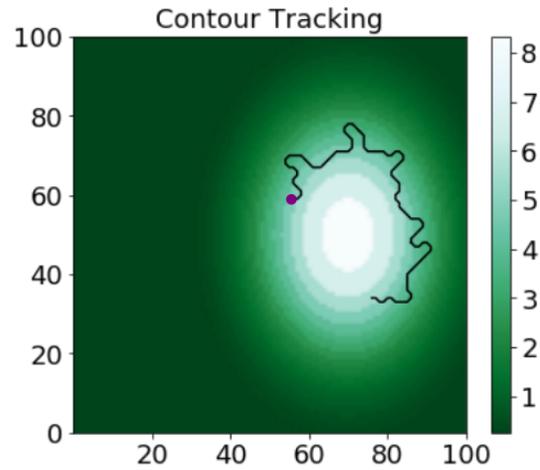

Fig.{6}. *The worm bot traces the contour around the setpoint concentration $C_T$ = 4.0Hz with the initial point having the same concentration as the setpoint*

In Fig.{5}, the initiating point of the simulation is at the purple dot with concentration value C = 2.56Hz and $C_T$ being the fixed 4.0Hz. The bot starts to ascend the gradient in a more defined trajectory than that of Fig.{4}, although it is not exactly a straight line due to the plateaus formed by quantization. Fig.{6} displays a typical contour tracking trajectory traced by the worm.

The absolute tracking deviation values are calculated by determining the maximum physical distance of the trajectory from the ideal contour of setpoint concentration 4Hz. We chose to calculate the deviation in this manner because deviation calculation solely based on concentration values will also capture the effects of non-linear encoding and quantization which is undesirable in this case.

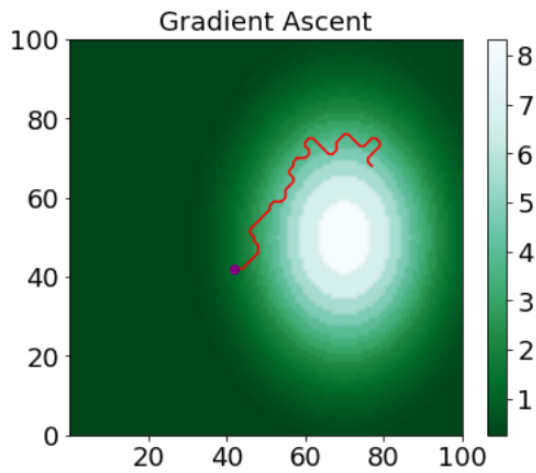

Fig.{5}. *The worm bot performs gradient ascent initiating from the purple dot with C = 2.56Hz and the specified setpoint being $C_T$ = 4.0Hz. Towards the end, it starts to perform contour tracking about the setpoint*

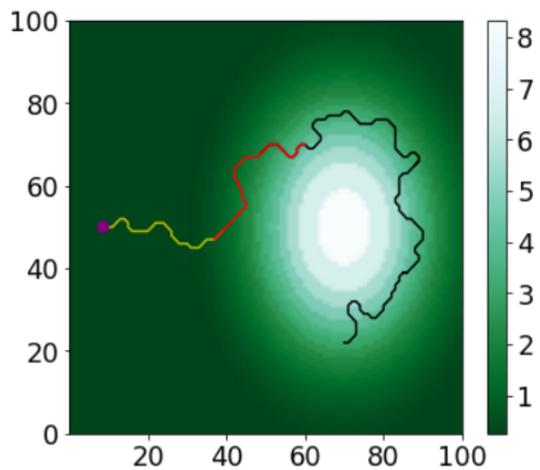

Fig.{7}. *The complete worm bot trajectory highlighting three sections - random walk (yellow), gradient ascent (red), and contour tracking (black)*

The complete Contour Tracking trajectory of the worm from the initial point (10,50) is plotted in Fig.{7} and the trajectory of the worm in the vicinity of a noisy concentration space with White Gaussian Noise of µ= 0 and

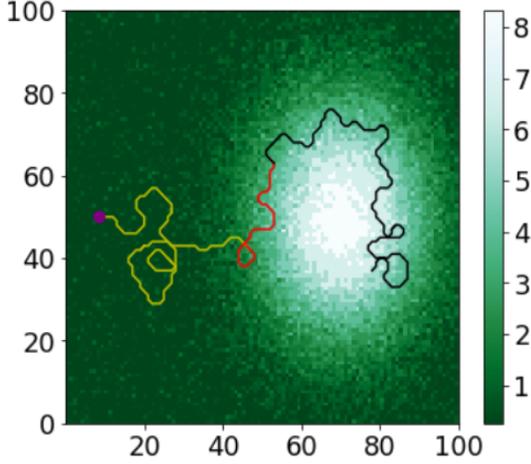

*Fig.{8}. The complete worm bot trajectory highlighting three sections - random walk (yellow), gradient ascent (red), and contour tracking (black) for a noisy concentration space with WGN of mean = 0 and standard deviation = 0.05*

We further implemented the same algorithm on software using Python, capturing the functionality of Interactive Spike Sender as well, to validate the correctness of operation and comparability of time consumption with Loihi. We ran a total of 40 simulations with the same initial conditions, random seed(different for subsequent simulations), and concentration setpoint in each simulation for Python and Loihi. The trajectories for both Python and Loihi are plotted for one of the simulations in Fig.{9}.

The simulation results for all three implementations are tabulated in Table.{3}. Setpoint foraging time is determined by measuring the time it takes for the worm to reach the setpoint for the first time during the simulation. The foraging time of Loihi's worm on average was found to be 1.92% greater than the foraging time of Python's worm which exhibits the time comparability of the two implementations.

| Simulation | Avg setpoint foraging time | Avg relative tracking deviation w.r.t Python | Avg difference in trajectory w.r.t Python |
|---|---|---|---|
| Loihi | 3906.13s | 1.216 | 6.73% |
| Loihi (Noisy) | 5689.36s | 1.414 | - |
| Python | 3832.42s | 1 | - |

*Table.{3}. Average setpoint foraging time, average contour tracking deviation w.r.t python and the average difference in full trajectory w.r.t python for Loihi (Noiseless), Loihi (Noisy), and Python implementations*

The table also shows the average contour tracking deviation for all the cases relative to the deviation observed in python. The difference in trajectory illustrates the closeness of the trajectories traced by the worm under the same conditions and parameters in Loihi w.r.t the trajectory traced by python. This difference arises because Loihi and Interactive Spike Sender have no common notion of time, which leads to small spike-time irregularities. These irregularities accumulate over time to produce more differences as the worm travels further. Hence we find that the normalized distance between the final points in the trajectories of Loihi and python exhibits the cumulative irregularities and is a good metric to represent the difference in their trajectories. This average difference of trajectories between Loihi and python is 6.73% which establishes the congruency in their trajectories.

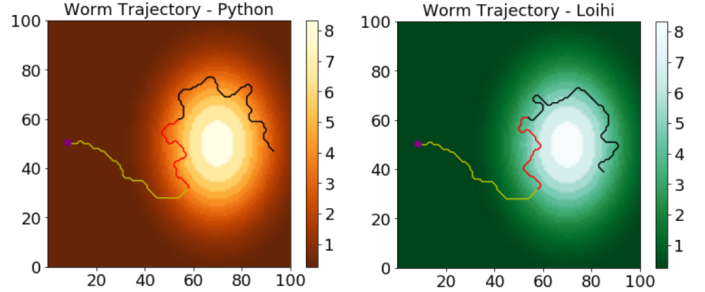

*Fig.{9}. The python simulation (left) of hardware friendly klinokinesis algorithm with the same set of parameters and random seed as of the Loihi simulation (Right). The initial point of simulation is (10,50), $C_T$ = 4.0Hz, sections of trajectory - random walk (yellow), gradient ascent (red), and contour tracking (black)*

Research has shown that Neuromorphic hardware reduces the simulation time (by ~2% in CPUs) and power consumption(by up to 100 times) [9][10] as compared to their software counterparts. We believe that the timings here are comparable and not definitive of the fastness/slowness of Loihi as compared to software because of the very small network used. These results hence highlight the fidelity, correctness, and time comparability of the hardware implementation of this algorithm with respect to the software.

### IV. CONCLUSION

In this paper, we showed that a simplified novel SNN architecture with only LIF neurons can be designed to translate a widely used complex spiking algorithm into a hardware-friendly algorithm. A purely LIF neuron-based design successfully eliminated external operations by approximating the Heaviside function and subtraction operation through spikes as well as replicating the outcome of the complex dynamics of ASE neurons. The algorithm is successfully implemented on Intel's Neuromorphic Processor Loihi to perform Contour Tracking with a very high correlation with its software counterpart. We further highlighted the robustness of the tracking algorithm by simulating the worm bot in a noisy environment. Thus, the entire signal processing of chemotaxis using klinokinesis was adapted into a completely spiking implementation with LIF neurons and demonstrated on Loihi. Thus, a critical control task is demonstrated on a generic SNN hardware - which can be integrated with other SNN blocks available on

Loihi like SLAM, learning recognition, path planning, etc towards energy-efficient SNN based autonomous bots.